\documentclass[onecolumn]{aa}
\usepackage[english]{babel}
\usepackage{graphicx}
\usepackage{amsmath}
\usepackage{array}
\usepackage{epsfig}
\usepackage{subfigure}
\usepackage{times}
\usepackage{mathptm}
\usepackage[latin1]{inputenc}
\usepackage{natbib}
\usepackage{longtable}
\begin{document}
\hyphenation{li-te-ra-tu-re va-ria-bi-li-ty}
\hyphenation{com-pa-ri-son re-pre-sent}

\title{REM near-IR and optical multiband observations of PKS\,2155-304 in
  2005\thanks{This paper is the corrected version astroph 0704.0265 published in
A\&A 469 503. It contains the material in the "Errata Corrrige", in press
in A\&A.}}

\author{A. Dolcini\inst{1}
 \and F. Farfanelli\inst{2}
 \and S. Ciprini\inst{2}
 \and A. Treves\inst{1}
 \and S. Covino\inst{3}
 \and G. Tosti\inst{2}
 \and E. Pian\inst{4}
 \and B. Sbarufatti\inst{1}
 \and E. Molinari\inst{3}
 \and G. Chincarini\inst{3}$^{,}$\inst{5}
 \and F.~M. Zerbi\inst{3}
 \and G. Malaspina\inst{3}
 \and P. Conconi\inst{3}
 \and L. Nicastro\inst{6}
 \and E. Palazzi\inst{6}
 \and V. Testa\inst{7}
 \and F. Vitali\inst{7}
 \and L.~A. Antonelli\inst{7}
 \and J. Danziger\inst{4}
 \and G. Tagliaferri\inst{3}
 \and E. Meurs\inst{8}
 \and S. Vergani\inst{8}
 \and A. Fernandez-Soto\inst{9}
 \and E. Distefano\inst{10}
 \and G. Cutispoto\inst{10}
 \and F. D'Alessio\inst{7}}

\institute{Universit$\mathrm{\grave{a}}$ degli Studi dell'Insubria, Dipartimento di Fisica e
 Matematica, via Valleggio 11, 22100 Como, Italy
 \and Dipartimento di Fisica e Osservatorio Astronomico,
 Universit$\mathrm{\grave{a}}$ di
 Perugia, Via. A. Pascoli, 06123 Perugia, Italy
 \and INAF, Osservatorio Astronomico di Brera, via E. Bianchi 46, 23807 Merate
 (LC), Italy
 \and INAF, Osservatorio Astronomico di Trieste, Via G. B. Tiepolo 11, 34143
 Trieste, Italy
 \and Universit$\mathrm{\grave{a}}$ degli Studi di Milano-Bicocca, Dipartimento di Fisica,
 Piazza delle Scienze, 3, 20126 Milan, Italy
 \and INAF/IASF Bologna, via Gobetti 101, 40129 Bologna, Italy
 \and INAF, Osservatorio Astronomico di Roma, via Frascati 33, 00040
 Monteporzio Catone, Italy
 \and Dunsink Observatory, Castleknock, Dublin 15, Ireland
 \and Observatori Astronomic, Universitat de Valencia, Aptdo. Correos 22085,
 Valencia 46071, Spain
 \and INAF, Osservatorio Astrofisico di Catania, via S. Sofia 78, 95123
 Catania, Italy}

\date{Received 29 September 2006 / Accepted 26 March 2007}

\abstract
{Spectral variability is the main tool for constraining emission models of BL
 Lac objects.}
{By means of systematic observations of the BL Lac prototype PKS 2155-304 in
 the infrared-optical band, we explore variability on the scales of months,
 days and hours.}
{We made our observations with the robotic 60 cm telescope REM located at La
 Silla, Chile. VRIJHK filters were used.}
{PKS 2155-304 was observed from May to
December 2005.
The wavelength interval explored, the
total number of photometric points and the short integration time render
our photometry substantially superior to
previous ones for this source. On the basis of the intensity and colour
we distinguish three different states of the source, each of duration of
months, which include all those described in the literature.
In particular, we report the highest state ever detected in the H band.
The source varied by a factor of 4 in this band, much more than in the
V band (a factor $\approx$ 2). The source softened with increasing intensity, contrary
to the general pattern observed in the UV-X-ray bands. On five nights of November
we had nearly continuous monitoring for 2-3 hours. A variability
episode with a time scale of  $\tau\approx$24 h is well documented, a much more rapid flare
with $\tau$=1-2 h, is also apparent, but is supported by relatively few points.}
{ The overall spectral energy distribution of PKS 2155-304 is commonly
   described by a synchrotron-self-Compton model. The optical infrared
   emission is however in excess of the expectation of the model, in its original
   formulation. This can be explained by a variation of the frequency of the
   synchrotron peak, which is not unprecedented in BL Lacs.}

\keywords{galaxies: active - galaxies: BL Lacertae objects: PKS 2155-304}

\authorrunning{A. Dolcini et al.}
\titlerunning{REM monitoring of PKS2155-304 during 2005}
\maketitle
\section{Introduction}

PKS 2155-304 (z=0.116, Falomo et al. 1991) is a prototype of high frequency peaked BL Lac objects. It has been
observed in the entire electromagnetic spectrum, from radio to TeV
gamma-rays. It was the target of several multifrequency campaigns,
the main scope of which was to study the variability of the spectral energy
distribution (SED), in order to constrain emission models.

In particular we
refer to the 1991 and 1994 campaigns involving IUE, ROSAT, ASCA, EUVE and
ground based telescopes (see Edelson et al. 1995, Urry et al. 1997, and
references therein). There were noticeable differences in source behaviour between these two epochs.
While in 1991 the multiwavelength variability was almost achromatic, and the X-ray
variation led that in the UV by a couple of hours, in 1994 the variability was
more pronounced in X-rays than in UV-optical, with a lag of the
latter by two days. The general pattern was that of a hardening of the spectrum
with increasing intensity. More recently Zhang et al. (2006b) studied a large set of
data covering the period 2000-2005 obtained with the XMM-Newton satellite, which
allowed a direct comparison of the X-ray and UV-optical band, the latter
deriving from the Optical Monitor on board the satellite. The complexity
of the variability pattern is confirmed. Some episodes of achromatic variation were
detected, but a general tendency of increasing variability amplitude with increasing
frequency, and spectral hardening with increasing intensity was found.

Optical photometry has been performed by several groups in several occasions (see e.g. Miller et al. (1983), Smith et al. (1992), Xie et al. (1996), Paltani et al. (1997), Pesce et
al. (1997), Fan \& Lin (2000), Tommasi et al. (2001) and references therein). All this
material is rather fragmented, consisting of few hours of observations during
few nights. The difficulty of a systematic observing campaign covering many
nights is partly
overcome by the possibility of observing using remotely guided or robotic
telescopes.

The REM telescope, originally designed for a prompt detection of
gamma ray bursts (see Molinari et al. (2006)), is particularly apt for photometric studies of BL Lacs
(see also the previous results for PKS 0537-441 by Dolcini et al. 2005, and
for 3C 454.3 by Fuhrmann et al. 2006)
and, being located at La Silla (Chile), it is ideally fit to study PKS 2155-304.

We report on extensive and intensive photometric campaign performed in 2005 in the V,
R, I, J, H, K bands. For the total number of photometric points, for the time
resolution (minutes) and spectral range this campaign seems to supersede
all the IR-optical photometric material presented thus far.

\section{REM, Photometric procedure, data analysis}

\subsection{REM}
The Rapid Eye Mount (REM) Telescope is a 60 cm fully robotic instrument. It has two cameras fed at the same time
by a dichroic filter that allows the telescope to observe in the NIR (z', J,
H, K) as well as
optical (I, R, V). Further information on the REM project may be found in Zerbi et al. (2001),
Chincarini et al. (2003) and Covino et al. (2004).

\subsection{Observations and data analysis}

REM observed the PKS 2155-304 field during May, September, October, November and
December 2005 in VRIH bands. Only during three nights in September the
telescope observed also in J and K filters. To allow intranight and
short time-scale variability monitoring, very intensive observations (2-3 h,
quasi-continuously) were made during five of the nights in November. An
outline of the observations is reported in Table 1, while the complete log is
only available in Table \ref{electronic_table} (see Appendix A): we report for each
photometric point the band, the epoch, the integration time, the intensity and its uncertainty. Typical
integration times are $\leq$100 s and statistical uncertainties are always $\leq$ 10$\%$ and $\leq$ 3$\%$ in the highest state
 (November 2005, see following).
\begin{table}
\begin{center}
\begin{tabular}{|c|c|c|c|}
\hline
\textbf{Period of observation} & \textbf{Nights of observation} &
\textbf{Number of photometric points} & \textbf{Total exposure time} \\
\hline
May & 6 & 129 & 14520 s \\
\hline
September & 8 & 159 &  18080 s \\
\hline
October & 3 & 102 & 11590 s \\
\hline
November & 21 & 1581 & 173540 s \\
\hline
December & 6 & 64 & 7030 s \\
\hline
\end{tabular}
\caption{Outline of observations accomplished in 2005.}
\label{outline}
\end{center}
\end{table}

\begin{figure}

\begin{center}

\includegraphics[width=8cm]{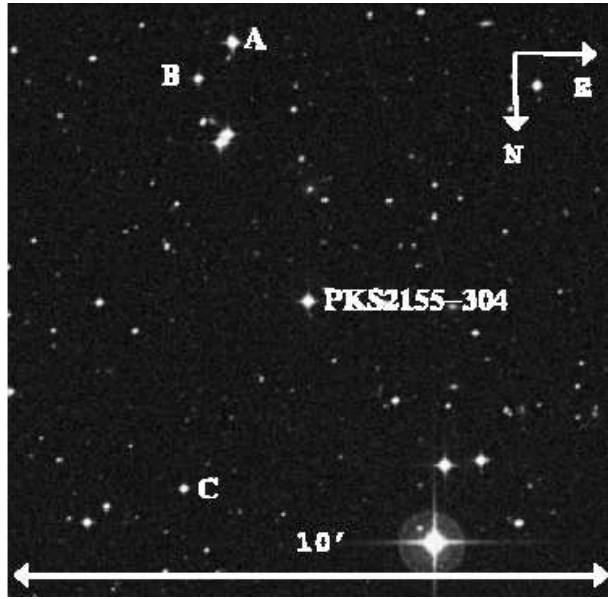}

\caption{PKS2155-304 field (DSS-1 survey). Letters indicate stars used for calibration.}

\label{frame}

\end{center}

\end{figure}

Reduction of the REM NIR and optical frames followed standard procedures.
Photometric analysis of the frames was done using the GAIA\footnote{http://star-www.dur.ac.uk/~pdraper/gaia/gaia.html} and DAOPHOT
packages (Stetson 1986).
Relative calibration was obtained by calculating magnitude shifts relative
to three bright isolated stars in the field, indicated by A, B, C in
Fig. \ref{frame} (image taken from ESO Digitized Sky Survey\footnote{http://archive.eso.org/dss/dss}).

The NIR frames were calibrated using the magnitudes of the A, B and C stars
as reported in the 2MASS catalogue\footnote{http://irsa.ipac.caltech.edu}. For
the optical, we exposed on 2006 June 29 the standard field
G156-31 (Landolt, 1992), and immediately after this the PKS 2155-304 field. We calculated the
zero points which were then used to calibrate all of our data. The observed magnitudes
in the REM filters for the reference objects A, B, and C are reported in
Table \ref{magabs}. We have monitored the relative intensities of the A, B, C
reference stars during the entire observation period, and we have detected no
indication of variability within 0.1 mag (error on the average $\leq$0.01
 mag).

\begin{table}

\begin{center}

\begin{tabular}{|c|c|c|c|c|}

\hline

\textbf{} & \textbf{A} & \textbf{B} & \textbf{C} \\
\hline
RA & 21:58:46.505 & 21:58:43.807 & 21:58:42.337 \\
\hline
DEC & -30:17:51.29 & -30:17:15.71 & -30:10:27.41 \\
\hline
K & 11.171$\pm$0.024 & 12.475$\pm$0.030 & 12.648$\pm$0.024 \\
\hline
H & 11.182$\pm$0.027 & 12.556$\pm$0.026 & 12.769$\pm$0.027 \\
\hline
J & 11.510$\pm$0.027 & 12.838$\pm$0.026 & 13.091$\pm$0.029 \\
\hline
I & 12.184$\pm$0.005 & 13.421$\pm$0.009 & 13.216$\pm$0.006 \\
\hline
R & 12.981$\pm$0.004 & 13.434$\pm$0.006 & 13.671$\pm$0.010 \\
\hline
V & 13.179$\pm$0.005 & 13.822$\pm$0.009 & 13.899$\pm$0.013 \\
\hline
\end{tabular}
\caption{Coordinates, IR and optical magnitudes for the reference stars.}
\label{magabs}
\end{center}
\end{table}

Note that we found significant
deviations from the optical calibrations provided by the finding charts for AGN of the Heidelberg
University\footnote{http://www.lsw.uni-heidelberg.de/projects/extragalactic/charts/2155-304.html}
(Hamuy \& Maza, 1989). In particular the star C is also used as
 a calibrator by these authors and our optical zeropoint differs by  about 0.3 mag
 from theirs.

 Relative and absolute calibration errors have been
added in quadrature to the photometric error derived from the procedure.
\section{Results}
\subsection{Long term variability}

\begin{figure}

\begin{center}

\includegraphics[width=12cm]{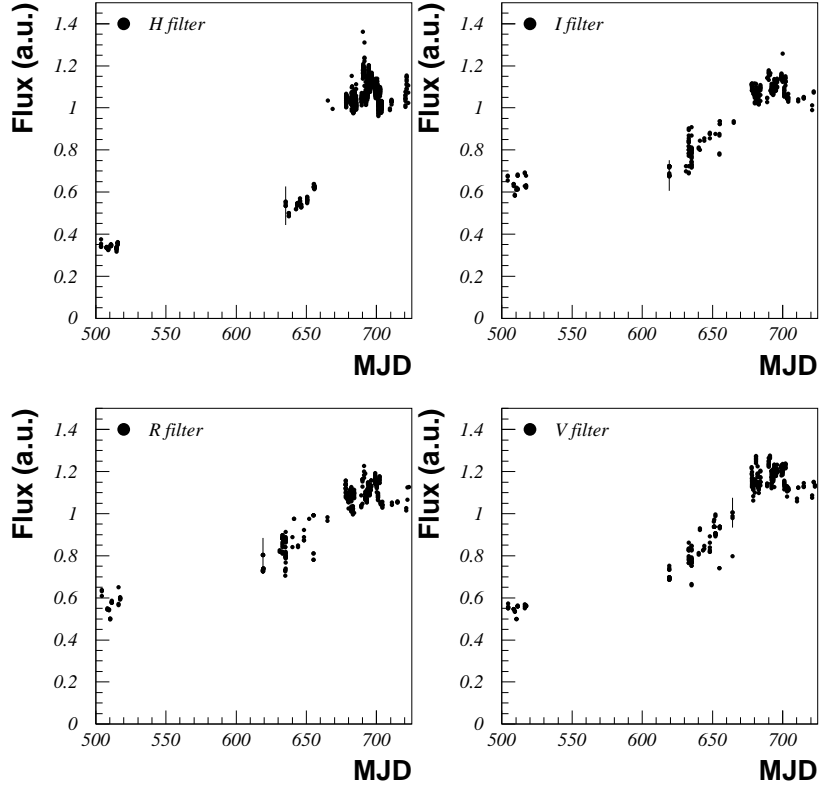}
\caption{Normalized light curves of PKS 2155-304. Flux
 is reported in arbitrary unit (a. u.). In each boxes a typical error bar
 is plotted.}
\label{curve_luce}
\end{center}
\end{figure}

In this section we report the results of the long term photometric
analysis. The light curves in the H, R, I, V filters are given in Fig. \ref{curve_luce}.

\begin{figure}

\begin{center}

\includegraphics[width=8cm]{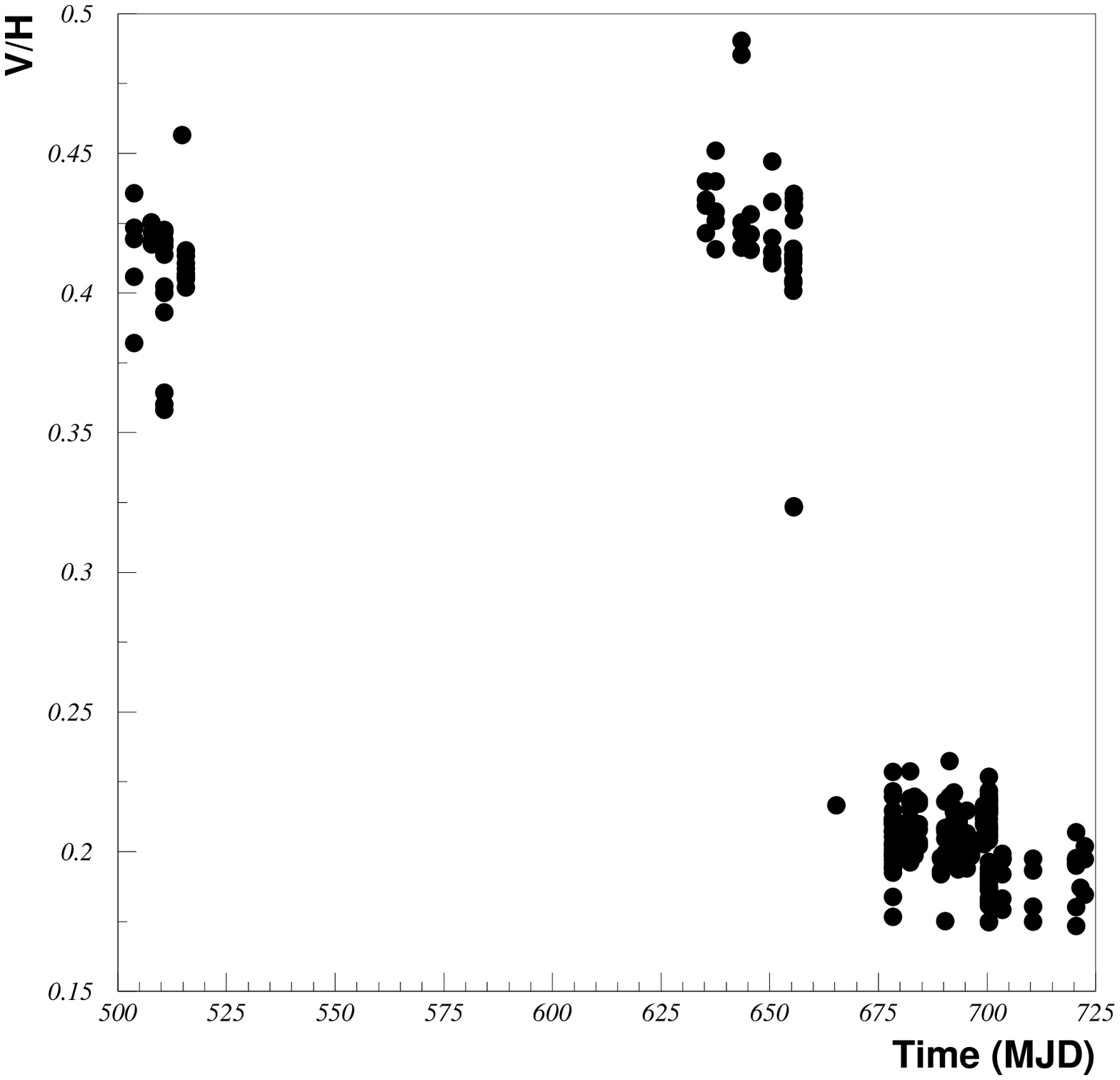}

\caption{V/H flux ratio evolution during 2005. Error bars are
   comparable with symbol size.}

\label{vsuhtime}

\end{center}

\end{figure}

The intensity is normalized
with respect to the average over the entire observation period. These
averages are given in Table \ref{table1}. It is immediately apparent that the total
variability range is very different in the various filters, being a factor
$\approx$ 4 in H and a factor $\approx$ 2 in V (see Table \ref{table1}) .
\begin{table}

\begin{center}

\begin{tabular}{|r|r|r|r|r|}

\hline

{\bf Filter} & {\bf H} & {\bf I} & {\bf R} & {\bf V} \\
\hline
{\bf Average} & 114.9$\pm$3.3 & 34.45$\pm$6.5 & 30.89$\pm$5.13 &
30.70$\pm$5.05 \\
\hline
{\bf Max value} & 156.5 & 46.4 & 38.3 & 37.4 \\
\hline
{\bf Min value} & 36.5 & 19.1 & 16.2 & 16.2 \\
\hline
{\bf Average ep.1} & 39.3$\pm$1.4 & 21.4$\pm$1.5 & 18.7$\pm$1.3 & 18.1$\pm$0.7
\\
\hline
{\bf Average ep.2} & 65.9$\pm$5.2 & 28.4$\pm$3.3 & 27.2$\pm$2.5 &
20.3$\pm$3.4\\
\hline
{\bf Average ep.3} & 122.9$\pm$6.1 & 38.8$\pm$1.9 & 34.1$\pm$1.5 & 33.5$\pm$1.7\\
\hline
\end{tabular}

\caption{Average intensities for all epochs and all filters. All data are in mJy
 units. \textbf{Epoch 1} corresponds to May 2005 observations, \textbf{epoch 2} to
 September-October 2005 observations and \textbf{epoch 3} to November-December 2005 observations.}

\label{table1}

\end{center}

\end{table}
The shapes of the light curves
are similar in the various filters. A flare-like structure is apparent in
all filters at t $\approx$ 680 (first days of November).
The ratio between the V- and H-band fluxes, designated as V/H, is reported in
Fig. \ref{vsuhtime}. In order not to introduce spurious effects due to small time scale
variability, the V/H ratio has been computed for pairs of V and H
measurements spaced apart in time by no more than 10 minutes.

 It seems that there
are two main colour states: the source softens rather abruptly, in response to the November flare. On the basis of the
light curve and the colour curve we divide the observations in three epochs:
\textbf{1} 500-525, \textbf{2} 640-660, \textbf{3} 670-725, expressed in
MJD\footnote{For the Modified Julian Date we use the convention
 MJD=JD-2,453,000.5}.

\subsection{Short time-scale variability}
\label{short}
\begin{figure}

\begin{center}

\includegraphics[width=12cm]{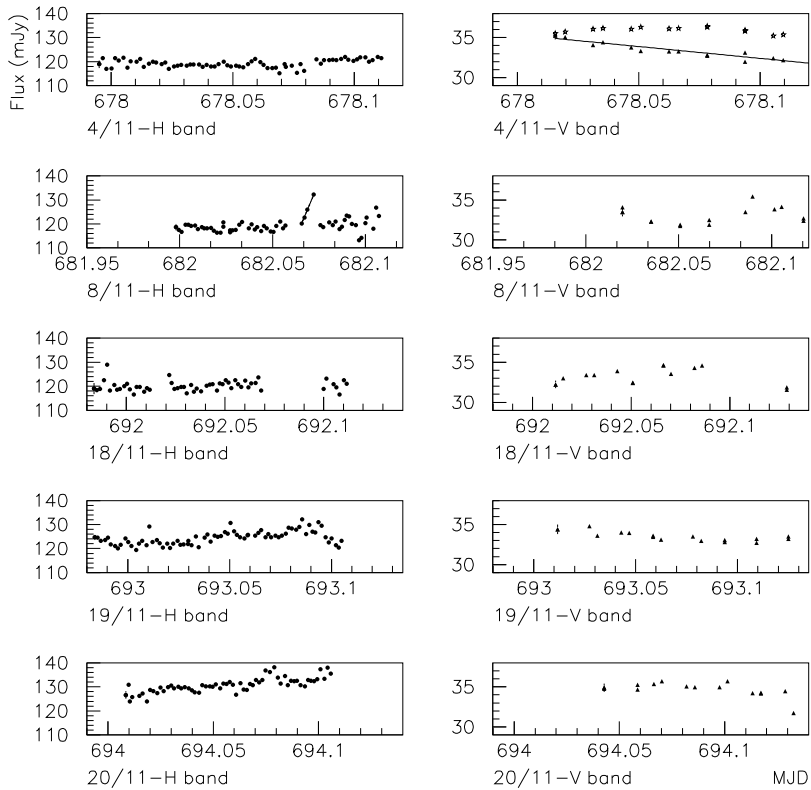}

\caption{Light curves in the H and V filters for five nigths in November 2005,
 when the observations were more intensive. Dates of observations are reported
in each box. The solid line in V band - 4 Nov box results from a linear
regression analysis. The solid line in H band - 8 Nov box connects the four
points of the flare-like structure. In each box it is given a typical
 error bar. In V band - 4 Nov box the
 light curve of one comparison star is also plotted, with a fixed enhancement of 9 mJy.}

\label{curvenov}

\end{center}

\end{figure}

We report in Fig. \ref{curvenov} the light curves for five nights in
November 2005, when
the observations were more intensive. All the
nights belong to epoch \textbf{3}, corresponding to the high state of the source.

\begin{figure}

\begin{center}

       \includegraphics[width=12cm]{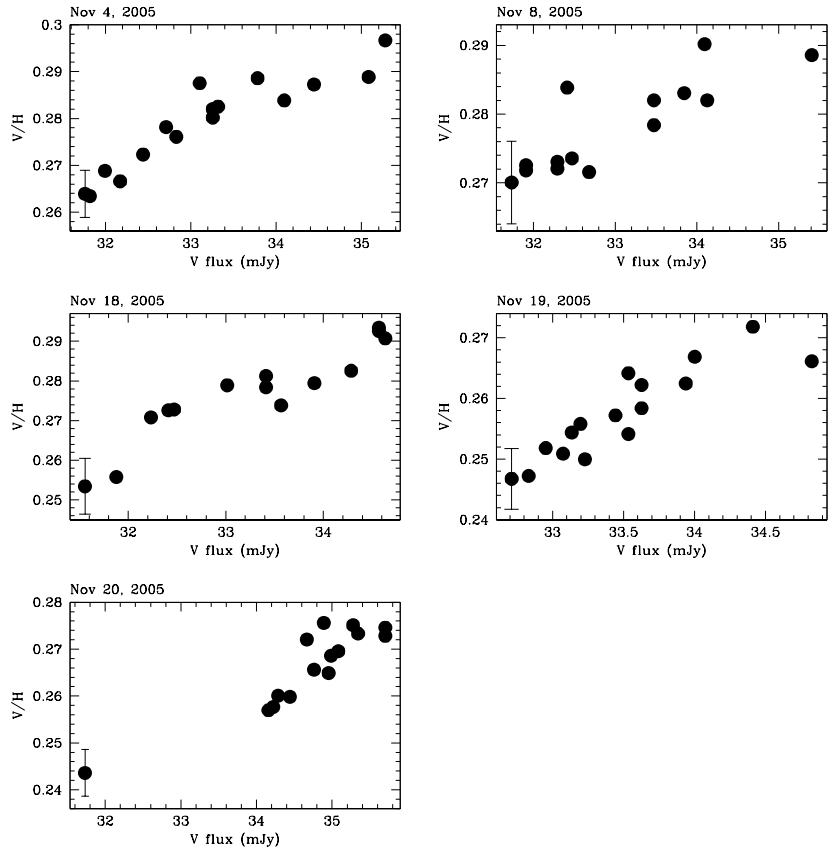}

       \caption{V/H flux ratio versus intensity for the five more intensively
         observed nights of epoch \textbf{3}. In each box a typical
         error bar is plotted.}

       \label{vsuhsuv}

       \end{center}

\end{figure}

The mean intensity and the 1-sigma values for each night are given in Table \ref{table4}.

\begin{table}

\begin{center}

\begin{tabular}{|r|r|r|r|r|r|}

\hline

{\bf Night} & {\bf 4/11} & {\bf 8/11} & {\bf 18/11} & {\bf 19/11} & {\bf 20/11} \\
\hline
{\bf Average H} & 119.3$\pm$1.7 & 119.3$\pm$3.0 & 120.4$\pm$2.1 & 124.5$\pm$2.8 &
130.8$\pm$3.0 \\
\hline
{\bf Average I} & 39.1$\pm$0.6 & 36.4$\pm$0.5 & 38.7$\pm$0.6 & 38.7$\pm$0.6 &
38.3$\pm$0.7 \\
\hline
{\bf Average R} & 38.5$\pm$0.8 & 36.40$\pm$.8 & 37.3$\pm$0.5 & 38.0$\pm$1.6 &
37.1$\pm$0.5 \\
\hline
{\bf Average V} & 33.2$\pm$1.1 & 33.0$\pm$1.1 & 33.4$\pm$1.1 & 33.5$\pm$1.1 &
34.7$\pm$0.1 \\
\hline

\end{tabular}

\caption{Average intensities and 1-sigma values for all filters for all five nights with more intensive observations in November 2005. All values are in mJy units.}

\label{table4}

\end{center}

\end{table}
A $\chi^{2}$ analysis indicates that in each night the significance of
 variability is very high, but for the nights of Nov 4 and Nov 18 for the
 H band and Nov 19 for the V band. In the box of Nov 4 - V band we also
 report the photometry of a comparison star which illustrates
 directly the significance of the source variability.
Though the shapes of intensity curves are different (see Fig. \ref{curvenov}), there is a  rather regular colour-intensity dependence (see Fig. \ref{vsuhsuv})
indicating harder states for higher intensities.

We adopt the usual definition of time scale variability
$\tau= \frac{1}{1+z} \frac{<f>} {df/dt}$. Following Montagni et
al. (2006), a variability time scale is taken as reliable
if the light curve can be approximated with a linear dependence, and it
contains at least 10 points. In particular this gives a time scale
of $\approx24$ h for the November 4 night (Fig. \ref{curvenov}, V band - Nov 4
box). The simultaneous H light curve does not show any regular variability. We note that on
November 8 in the H curve there is
a flare-like event. If one connects 4 points as suggested in
Fig. \ref{curvenov} H band - Nov 8 box,
the time scale variability is as short as 1.5 h. Unfortunately the V light curve is
too sparse to confirm the presence of the flare also in this band.

\subsection{The NIR-Optical spectral energy distribution}

\begin{figure}

\begin{center}

\includegraphics[width=8cm]{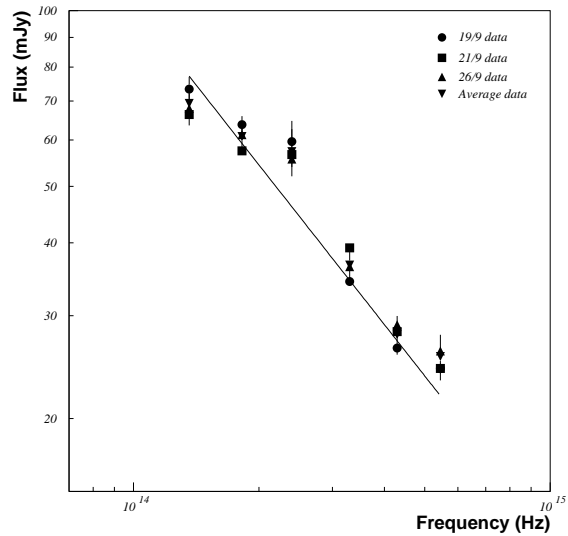}

\caption{September 2005 spectra for observations including the K and J filters.
 The spectral fit on average data with a single power law yields a
 spectral index $\alpha$=0.91$\pm$0.07.}

\label{spettri_set}

\end{center}

\end{figure}

We had six filter coverage (K,H,J,I,R,V) during three nights
of Sept. 2005 (epoch \textbf{2}) and representative SEDs for these nights are reported in Fig. \ref{spettri_set}.

The delays between exposures in the different filters are less than 20
minutes. Reddening corrections are less than 6\% in V and have been neglected.
A fit with a single power law yields $\alpha\approx$0.9 and it is clearly not
good. The main deviation derives from the J filter, exceeding substantially
our photometric precision of about 10\%.
An improvement in the fit is obtained by using a broken power law with
spectral indices $\alpha\approx$0.4 for the IR data and $\alpha\approx$0.9 for
the optical data.

\begin{figure}

\begin{center}

\includegraphics[width=8cm]{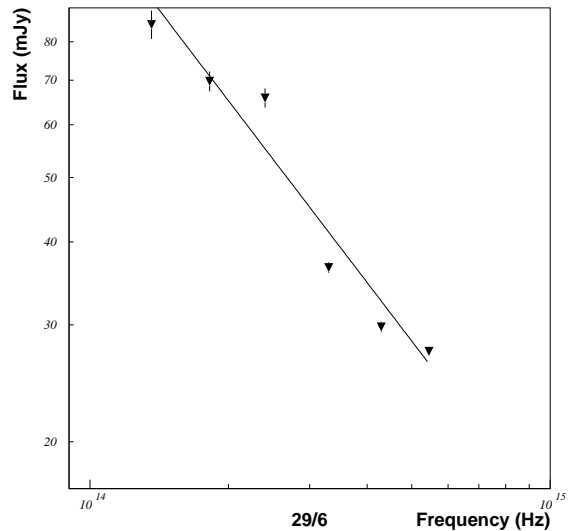}

\caption{29 June 2006 spectrum. The spectral fit with a single power
law yields a spectral index $\alpha$=0.90$\pm$0.16.}

\label{spettro29giu}

\end{center}

\end{figure}

For comparison, we report in Fig. \ref{spettro29giu} the SED of June 29, 2006,
exposure used for calibration purpose: its profile is rather similar to that of Sept. 2005.

\begin{figure}

\begin{center}

\includegraphics[width=8cm]{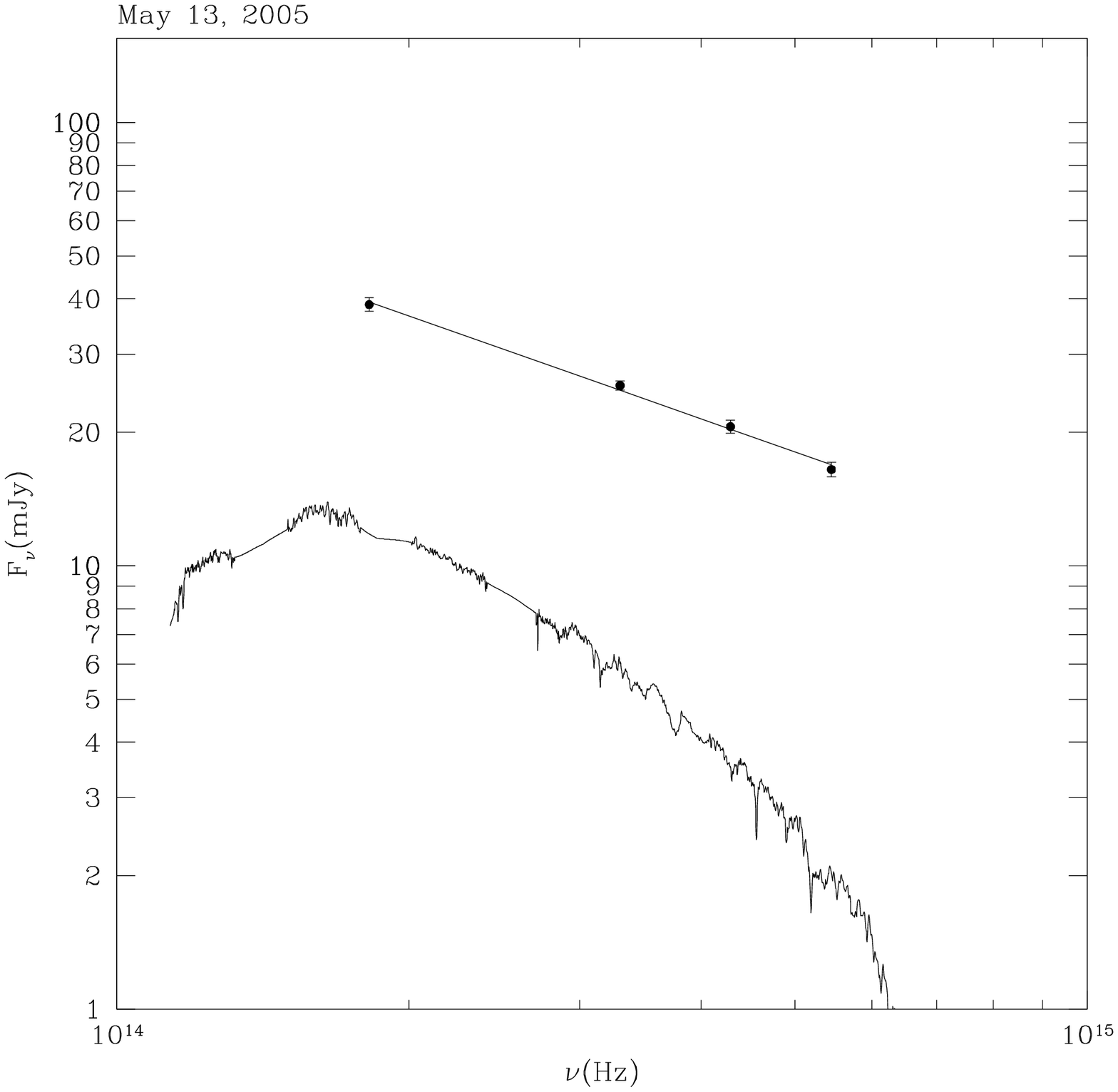}

\caption{13 May 2005 spectrum - epoch 1. We report also the
 spectrum of the host galaxy  (see text). The spectral fit with a single power law
 yields a spectral index $\alpha$=0.77$\pm$0.16.}

\label{spettro13mag}

\end{center}

\end{figure}

At the other epochs the SED consists of 4 points (H, I, R, V), and in
Figs. \ref{spettro13mag} and \ref{spettro4nov}
we give representative examples of SEDs acquired on epoch \textbf{1} and \textbf{3}. The
time differences between observations at various filters are less than 20 minutes.

\begin{figure}

\begin{center}

\includegraphics[width=8cm]{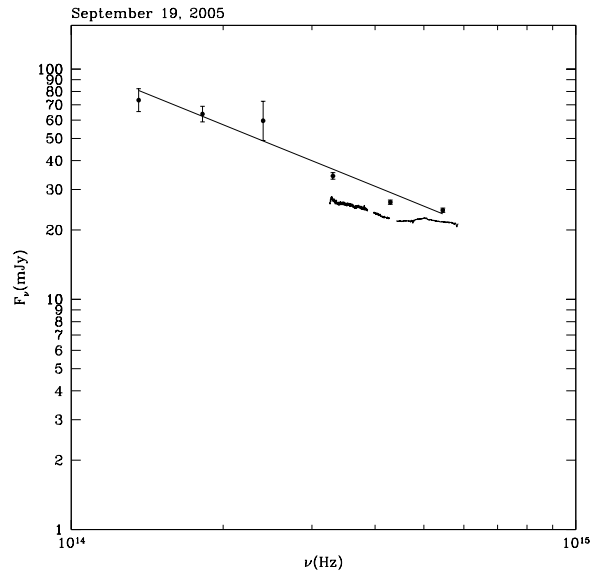}

\caption{19 September 2005 spectrum - epoch 2. For comparison we report the ESO 3.6m telescope
 spectrophotometry which correspond to a slightly lower state of the
 source. The spectral fit with a single power law yields a spectral index $\alpha$=0.88$\pm$0.05.}

\label{spettro19set}

\end{center}

\end{figure}

\begin{figure}

\begin{center}

\includegraphics[width=8cm]{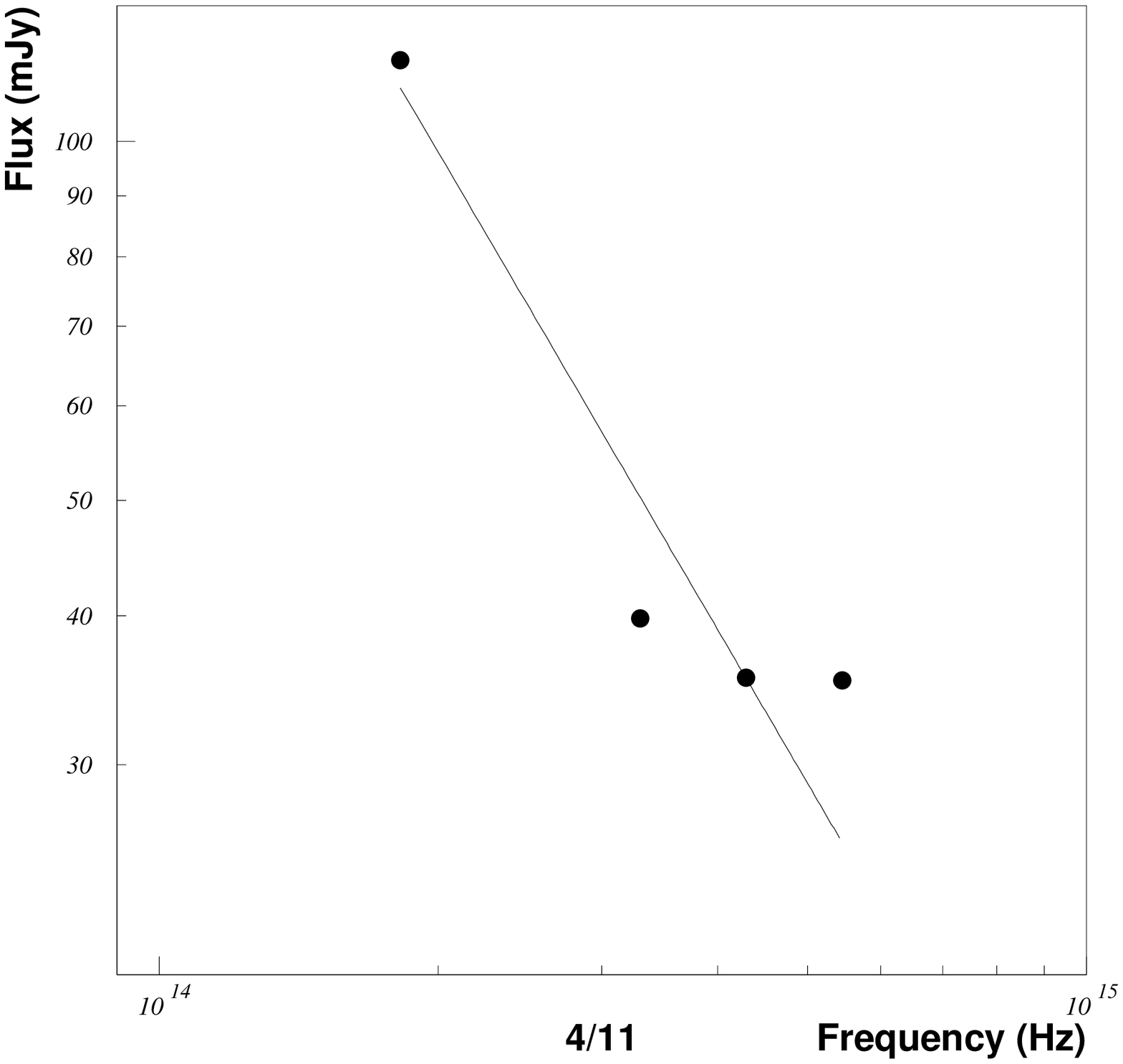}

\caption{4 November 2005 spectrum - epoch 3. The spectral fit with a
single power law yields a spectral index $\alpha$=1.32$\pm$0.25. Error bars are
comparable with symbols size.}

\label{spettro4nov}

\end{center}

\end{figure}

In Fig. \ref{spettro13mag}, which refers to a low state,
we report also the estimated contribution of the host galaxy, which was
calculated adopting the H magnitude of the galaxy measured by Kotilainen et
al. (1998) and the Mannucci at al. (2001) template spectrum for giant ellipticals. It is apparent that the
contribution of the galaxy never exceeds 20$\%$ of the BL Lac signal. At the other epochs
the contribution from galaxy is negligible and it is not relevant for
explaining the excess in J with respect a single power law noted above. The epoch \textbf{2} photometry (Fig. \ref{spettro19set})
is compared with spectrophotometry obtained with the ESO 3.6m telescope by R. Falomo\footnote{spectrum available at the ZBLLAC online library,
http://www.oapd.inaf.it/zbllac} on July 25, 2001 (Sbarufatti et al. (2006)). The source was found in a similar, but
somewhat lower brightness state and some deviations from a power law are apparent.
The HRIV points at epoch \textbf{3} (Fig. \ref{spettro4nov}) are roughly fitted by a single power law of
$\alpha\approx$1.3. In any case the comparison of the SEDs at the three epochs clearly indicate
a softening with increasing intensity.

\section{Discussion}

\begin{figure}

\begin{center}

\includegraphics[width=12cm]{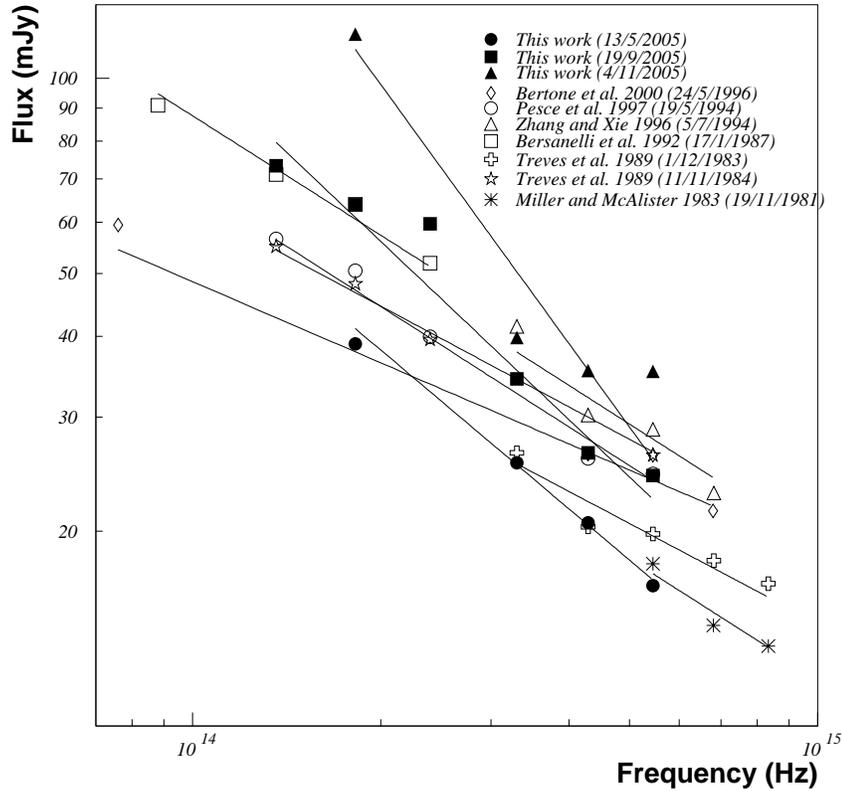}

\caption{Different spectra of PKS2155-304 from observations at other epochs reported in the
literature. Symbols correspond to following works: filled circles: this work (13/5/2005 data),
filled squares: this work (19/9/2005 data), filled up triangles: this work (4/11/2005
data), open diamonds: Bertone et al. (2000; 24/5/1996 data), open circles:
Pesce et al. (1997; 19/5/1994 data, the Hamuy \& Maza (1989) calibration is used), open up triangles: Zhang and Xie (1996;
5/7/1994 data), open squares: Bersanelli et al. (1992; 17/1/1987 data), open crosses:
Treves et al. (1989; 1/12/1983 data), open stars: Treves et
al. (1989; 11/11/1984 data), asterisks: Miller and McAlister (1983; 19/11/1981
data). Spectral
index values and V magnitudes for all data sets are reported in Table
\ref{indici_spettrali_V}.}

\label{spettri_storici}

\end{center}

\end{figure}

A collection of near-IR/optical SEDs of PKS2155-304 obtained by various authors at different epochs
is presented in Fig. \ref{spettri_storici} and in Table \ref{indici_spettrali_V}. Our data encompass all those reported in the literature.

\begin{table}

\begin{center}

\begin{tabular}{|c|c|c|}

\hline
\textbf{Data set} & \textbf{$\alpha$} & V (mJy) \\
\hline
This work (13/5/2005) & 0.77$\pm$0.16 & 16.485$\pm$0.263 \\
\hline
This work (19/9/2005) & 0.88$\pm$0.05 & 24.370$\pm$0.238 \\
\hline
This work (4/11/2005) & 1.32$\pm$0.24 & 35.278$\pm$0.498 \\
\hline
Bertone at al. (2000) & 0.42$\pm$0.26 & 26.20$\pm$0.58 \\
\hline
Pesce et al. (1997) & 0.62$\pm$0.30 & 24.50$\pm$0.67  \\
\hline
Zhang \& Xie (1996) & 0.62$\pm$0.16 & 22.90$\pm$0.63 \\
\hline
Bersanelli et al. (1992) & 0.61$\pm$0.38 & 51.88$\pm$1.56 (J band) \\
\hline
Treves et al. (1989) (1/12/1983) & 0.51$\pm$0.31 & 19.80$\pm$0.36 \\
\hline
Treves et al. (1989) (11/11/1984) & 0.51$\pm$0.41 & 26.20$\pm$0.48 \\
\hline
Miller \& McAlister (1983) & 0.62$\pm$0.56 & 17.8 \\
\hline
\end{tabular}
\caption{Spectral index values and V values for all spectra plotted in
 Fig. \ref{spettri_storici}. $\alpha$ vs V plot is reported in Fig. \ref{alpha_V}.}
\label{indici_spettrali_V}
\end{center}
\end{table}

\begin{figure}
\begin{center}
\includegraphics[width=10cm]{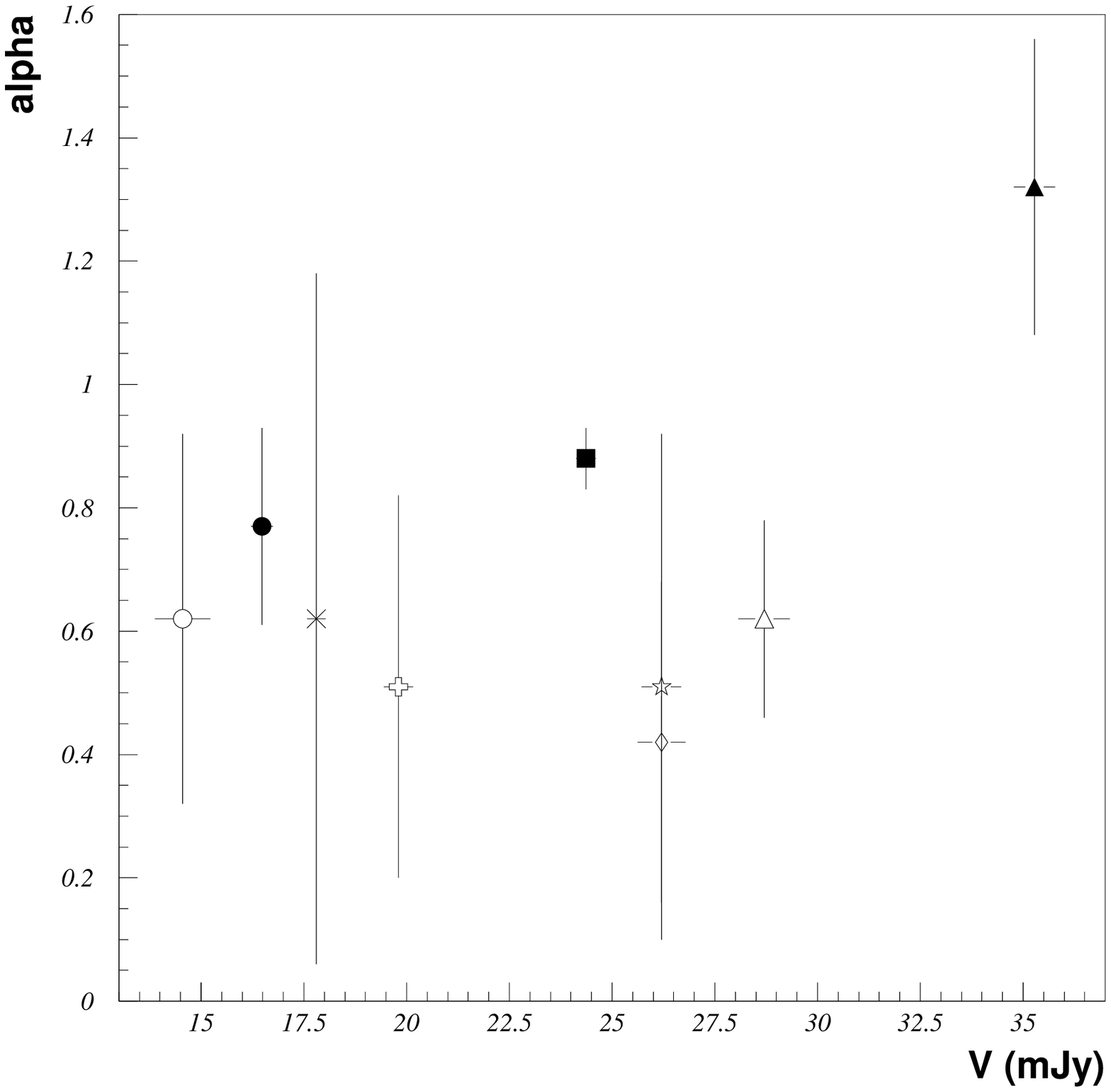}
\caption{$\alpha$ vs V plot for data reported in
 Fig. \ref{spettri_storici}. Symbols are the same as used in
 Fig. \ref{spettri_storici}.}
\label{alpha_V}
\end{center}
\end{figure}
In the historical observations of PKS2155-304 the delays between exposures
at different filters are typically of the order of hours, instead of about
10 minutes as in our data set.
Comparing literature data with our data it is apparent that the maximum
we observed on 20 November 2005 in the H filter light curve
is the highest state ever reported in this band. Note that the V state was
comparable with states reported in the literature, likely because the
coverage of
the source in the optical band is less sparse than that in the NIR.
A most noticeable result of our photometry is the discovery of long term
H-band variability, the amplitude of which is much larger than that in the
optical.

 In Fig. \ref{alpha_V} we plot the spectral index vs the V magnitude, as reported in table \ref{indici_spettrali_V}. There is no
 apparent correlation. It is noticeable however that the highest state in all
 bands (our observation of Nov 2005) corresponds to a rather soft spectral shape. This contrasts with the usual source behaviour of
hardening with increasing intensity, as found in the UV-X-ray band
(see Introduction).
It contrasts also with the short time scale variability, as reported
in section \ref{short}.

\begin{figure}
\begin{center}
\includegraphics[width=14cm]{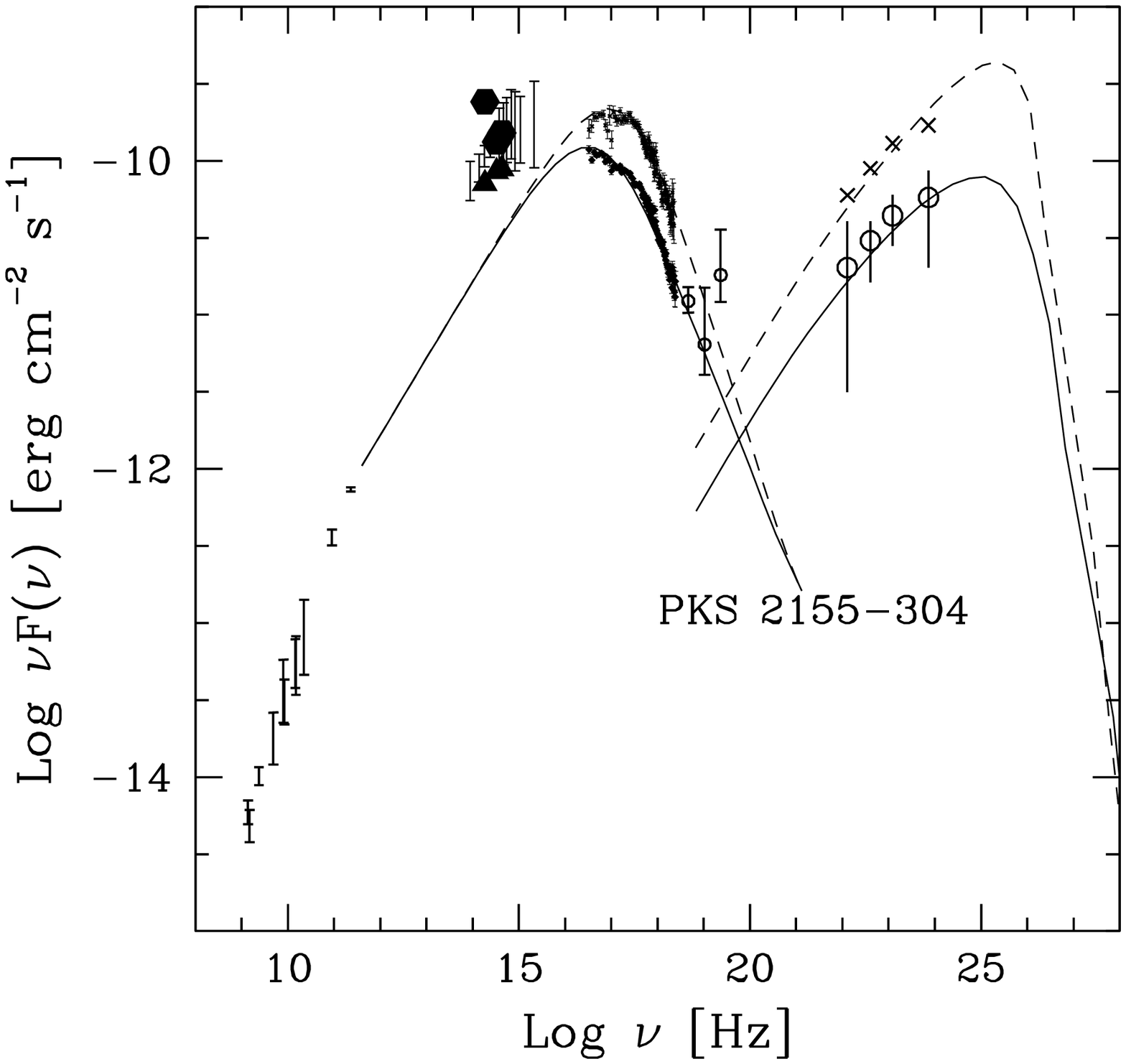}
\caption{ SED of PKS 2155-304 in two states, adapted from Chiappetti et
 al. (1999) (see the paper for details). Data from this work are
 also plotted. Filled triangles correspond to epoch \textbf{1} (13/5/2005
 data), while filled hexagons belong to epoch \textbf{3} data
 (20/11/2005). Optical, UV and REM data are dereddened using E(B-V)=0.026 and parameters given by Cardelli et
 al. (1989).}
\label{Chiappetti99}
\end{center}
\end{figure}

There is a general consensus that the blazar SED can be explained by
the superposition of a synchrotron component,
and an inverse Compton one due either to scattering off the synchrotron
photons (synchrotron-self Compton, SSC), or to external photons like those
of the broad line region or
of a thermal disk (e.g. Tavecchio et al. 1998, Katarzynski et al. 2005). This
results in a typical two-maxima shape of the
blazar SED. In Fig. \ref{Chiappetti99} we report examples of the  SED
modeling proposed for PKS 2155-304,
on the basis of data taken in 1997. The models are detailed in
 Chiappetti et al. (1999).
The object is a typical HBL, with the synchrotron peak in the soft
X-rays.

A well known critical point of this  model, is that the source
size is essentially constrained by variability, and variability itself
requires that the SED is constructed using simultaneous observations in
all bands. A further step of the modelling consists in identifying
the physical origin
of the relativistic jet and of its variability, see e.g. Katarzynski \&
Ghisellini (2006).  With this
premise it is obvious that the optical-IR photometric study, non simultaneous
with that in other regions of the SED, has only a limited relevance in
clarifying
the overall picture. However we would like to make some remarks.  If the
SSC models reported in Fig. \ref{Chiappetti99} truly represent the behaviour
of the SED in 1997, as suggested by the good match with the X-ray and TeV energy data, and if our 2005 optical-IR spectra are also due to the SSC mechanism, then the latter represent a
different condition in the jet and point to different critical parameters within the SSC scenario.
While the IR-optical spectrum in May 2005 (triangles) has the same shape as predicted in 1997, but different normalization, the November 2005 IR-optical spectrum is different in both shape and
normalization.  The May 2005 observation
suggests that the synchrotron peak may be located at a frequency
similar to the one observed in 1997 (approximately between extreme UV and soft X-rays), the
total energy being somewhat higher (about a factor 2, see Figure 13) than observed in 1997.  The
slope of the November 2005 spectrum suggests instead a  much lower synchrotron peak energy, around
the IR-optical domain or even redward, i.e. about 2-3 orders of magnitude lower than observed in 1997 and inferred in May 2005.  While a variation of the synchrotron peak energy of this amplitude and on this time scale
(the September 2005 slope is intermediate between those of May and November 2005, suggesting a
monotonic change) it is not unprecedented in blazars (Mkn501 exhibited a similar variation in a much
more rapid time scale, Pian et al. 1998),
this would be the first observation of this kind in PKS~2155-304.  Therefore, our interpretation is
only tentative, although supported by the large observed IR variability.


 Alternatively, in order to explain the optical-IR flux excess we observe in 2005 with respect to the
SSC prediction based on the earlier multiwavelength data (Fig. 13), one could
invoke a  thermal component, possibly from hot dust associated with the
  ``dusty torus'' surrounding the central region of the active nucleus, as suggested in the cases of other blazars with
excess in the optical-infrared band
(De Diego et al. 1997, for blazar 3C 66A; Pian et al. 1999 for 3C~279;
Pian et al. 2002, 2006, for blazar PKS 0537-441).  However, this seems
  somewhat less
likely, because high emission states, as observed by us, are expected to be
dominated by non-thermal beamed relativistic radiation.

 The continuation of this and other similar optical-IR studies, which have
  been proven to be promising but do not provide enough information for a
  physical interpretation of the data, requires that the observations are
  extended to other wavelengths. Simultaneous observations over a large
  wavelength range is the only tool to provide the necessary information for a
  physical interpretationof the observed variability of blazars.
REM
monitorings of the kind reported here
could be an effective trigger to X-ray satellites, and programs
along these lines are foreseen with SWIFT. Cross correlation procedures,
which up to now have been limited mainly to the X-ray band (Zhang et al. 2005,
2006a, 2006b, Sembay et al. 2002, Edelson et al. 1995), would be extended to
a much larger portion of the SED.

\appendix
\section{Table of observations}
\label{appendix}
\begin{longtable}{|c|c|c|c|c|}
\hline
\textbf{Filter} & \textbf{Epoch (MJD)}  &  \textbf{Integr time s}  &
\textbf{Intensity mJy}  &  \textbf{Sigma}  \\
\hline
K       &       633.4872        &       120     &       73.993  &       3.640   \\
K       &       633.4966        &       120     &       73.345  &       3.640   \\
K       &       633.5012        &       120     &       73.382  &       2.938   \\
K       &       633.5094        &       120     &       76.417  &       1.043   \\
K       &       633.5254        &       120     &       72.309  &       2.698   \\
K       &       635.4835        &       120     &       70.598  &       2.579   \\
K       &       635.6612        &       120     &       66.373  &       2.810   \\
\end{longtable}

\end{document}